\pdfoutput=1
\documentclass[conference,compsoc]{IEEEtran}
\ifCLASSOPTIONcompsoc
  \usepackage[nocompress]{cite}
\else
  \usepackage{cite}
\fi
\usepackage[pdftex]{graphicx}
\usepackage[caption=false]{subfig}
\usepackage{url}

\usepackage{mdwlist}
\usepackage{array}
\usepackage{enumitem}
\usepackage{multirow}
\usepackage[multiple]{footmisc}
\usepackage{amsmath}

\begin{document}

\title{Performance Overhead Comparison between Hypervisor and Container based Virtualization}

\author{\IEEEauthorblockN{Zheng Li\IEEEauthorrefmark{1},
Maria Kihl\IEEEauthorrefmark{1},
Qinghua Lu\IEEEauthorrefmark{2} and
Jens A. Andersson\IEEEauthorrefmark{1}}
\IEEEauthorblockA{\IEEEauthorrefmark{1}Department of Electrical and Information Technology, Lund University, Lund, Sweden\\
Email: \{zheng.li, maria.kihl, jens\_a.andersson\}@eit.lth.se}
\IEEEauthorblockA{\IEEEauthorrefmark{2}College of Computer and Communication Engineering, China University of Petroleum, Qingdao, China\\
Email: dr.qinghua.lu@gmail.com\\
}}

\maketitle

\begin{abstract}
The current virtualization solution in the Cloud widely relies on hypervisor-based technologies. Along with the recent popularity of Docker, the container-based virtualization starts receiving more attention for being a promising alternative. Since both of the virtualization solutions are not resource-free, their performance overheads would lead to negative impacts on the quality of Cloud services. To help fundamentally understand the performance difference between these two types of virtualization solutions, we use a physical machine with ``just-enough" resource as a baseline to investigate the performance overhead of a standalone Docker container against a standalone virtual machine (VM). With findings contrary to the related work, our evaluation results show that the virtualization's performance overhead could vary not only on a feature-by-feature basis but also on a job-to-job basis. Although the container-based solution is undoubtedly lightweight, the hypervisor-based technology does not come with higher performance overhead in every case. For example, Docker containers particularly exhibit lower QoS in terms of storage transaction speed. 
\end{abstract}


\IEEEpeerreviewmaketitle

\section{Introduction}
As a key element of Cloud computing, virtualization plays various vital roles in supporting Cloud services, ranging from resource isolation to resource provisioning. The existing virtualization technologies can roughly be distinguished between the hypervisor-based and the container-based solutions. Considering their own resource consumption, both virtualization solutions inevitably introduce performance overheads when offering Cloud services, and the performance overheads could then lead to negative impacts to the corresponding quality of service (QoS). Therefore, it would be crucial for both Cloud providers (e.g., for improving infrastructural efficiency) and consumers (e.g., for selecting services wisely) to understand to what extend a candidate virtualization solution incurs influence on the Cloud's QoS. 


Given the characteristics of these two virtualization solutions (cf.~the background in Section \ref{sec:relatedWork}), an intuitive hypothesis could be: \textit{a container-based service exhibits better performance than its corresponding hypervisor-based VM service}. Nevertheless, there is little quantitative evidence to help test this hypothesis in an ``apple-to-apple" manner, except for those qualitative discussions. Therefore, we decided to use a physical machine with ``just-enough" resource as a baseline to quantitatively investigate and compare the performance overheads between the container-based and hypervisor-based virtualizations. In particular, since Docker is currently the most popular container solution \cite{Pahl_2015} and VMWare is one of the leaders in the hypervisor market \cite{Walters_Chaudhary_2008}, we chose Docker and VMWare Workstation 12 Pro to represent the two virtualization solutions respectively.

According to the clarifications in \cite{Feitelson_2007}, our qualitative investigations can be regulated by the discipline of experimental computer science (ECS). By employing ECS's recently available Domain Knowledge-driven Methodology (DoKnowMe) \cite{Li_OBrien_DoKnowMe}, we experimentally explored the performance overheads of different virtualization solutions on a feature-by-feature basis. 

The experimental results and analyses 
show that the aforementioned hypothesis is not true in all the cases. For example, we do not observe computation performance difference between those service types with respect to solving a combinatorially hard chess problem; and the container even leads to higher storage performance overhead than the VM when reading/writing data byte by byte. Moreover, we find that the remarkable performance loss incurred by both virtualization solutions usually appears in the performance variability. Overall, the contributions of this work are threefold:

\begin{itemize}
	\item Our experimental results and analyses can help both researchers and practitioners to better understand the fundamental performance of the container-based and hypervisor-based virtualization technologies. It is notable that the performance engineering in ECS can roughly be distinguished between two stages: the first stage is to reveal the primary performance of specific (system) features, while the second stage is generally based on the first-stage evaluation to investigate real-world application cases. Thus, this work can be viewed as a foundation for more sophisticated evaluation studies in the future.  
	\item Our method of calculating performance overhead can easily be applied or adapted to different evaluation scenarios by others. The literature shows that the ``performance overhead" has normally been used in the context of qualitative discussions. By quantifying such an indicator, our study essentially provides a concrete lens into the case of performance comparisons.
	\item The whole evaluation logic and details reported in this paper can be viewed as a reusable and traceable template for evaluating Docker containers. Since the Docker project is still quickly growing \cite{Merkel_2014}, the evaluation results could gradually be out of date. Benefiting from this template, future evaluations can conveniently be repeated or replicated by different evaluators at different times and locations. 
\end{itemize}

The remainder of this paper is organized as follows. Section \ref{sec:relatedWork} summarizes the background knowledge of container-based and the hypervisor-based virtualizations and highlights the existing work related to their performance comparisons. Section \ref{sec:Implementation} introduces the performance evaluation implementation including the methodology employed in our study. The detailed performance overhead investigation is divided into two reporting parts, namely pre-experimental activities and experimental results \& analyses, and they are correspondingly described into Section \ref{subsec:preExperimental} and \ref{subsec:results} respectively. Conclusions and some future work are discussed in Section \ref{sec:conclusion}.

\section{Background and Related Work}
\label{sec:relatedWork}

\begin{figure}[!t]
  \centering
  \subfloat[Hypervisor-based virtual service.]{
    \label{fig:subfigHypervisor} 
    \includegraphics[width=3.9cm]{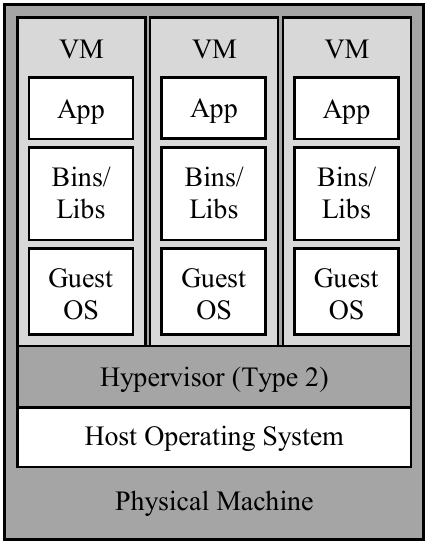}}
 \quad
  \subfloat[Container-based virtual service.]{
    \label{fig:subfigContainer} 
    \includegraphics[width=3.9cm]{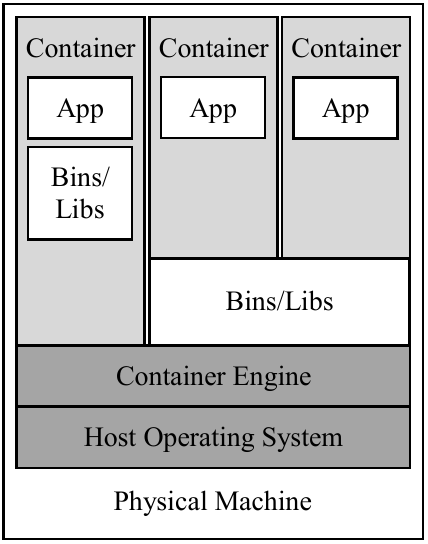}}
  \caption{Different architectures of hypervisor-based and container-based virtual services.}
  \label{fig:architecture} 
\end{figure}

When it comes to the Cloud virtualization, the de facto solution is to employ the hypervisor-based technologies, and the most representative Cloud service type is offering virtual machines (VMs) \cite{Xu_Yu_2014}. In this virtualization solution, the hypervisor manages physical computing resources and makes isolated slices of hardware available for creating VMs \cite{Merkel_2014}. We can further distinguish between two types of hypervisors, namely the bare-metal hypervisor that is installed directly onto the computing hardware, and the hosted hypervisor that requires a host operating system (OS). To make a better contrast between the hypervisor-related and container-related concepts, we particularly emphasize the second hypervisor type, as shown in Figure \ref{fig:subfigHypervisor}. Since the hypervisor-based virtualization provides access to physical hardware only, each VM needs a complete implementation of a guest OS including the binaries and libraries necessary for applications \cite{Bernstein_2014}. As a result, the guest OS will inevitably incur resource competition against the applications running on the VM service, and essentially downgrade the QoS from the application's perspective. 

To relieve the performance overhead of hypervisor-based virtualization, researchers and practitioners recently started promoting an alternative and lightweight solution, namely container-based virtualization. In fact, the foundation of the container technology can be traced back to the Unix \texttt{chroot} command in 1979 \cite{Bernstein_2014}, while this technology is eventually evolved into virtualization mechanisms like Linux VServer, OpenVZ and Linux Containers (LXC) along with the booming of Linux \cite{Xavier_Neves_2014}. Unlike the hardware-level solution of hypervisors, containers realize virtualization at the OS level and utilize isolated slices of the host OS to shield their contained applications \cite{Bernstein_2014}. In essence, a container is composed of one or more lightweight images, and each image is a prebaked and replaceable file system that includes necessary binaries, libraries or middlewares for running the application. In the case of multiple images, the read-only supporting file systems are stacked on top of each other to cater for the writable top-layer file system \cite{Pahl_2015}. With this mechanism, as shown in Figure \ref{fig:subfigContainer}, containers enable applications to share the same OS and even binaries/libraries when appropriate. As such, compared to VMs, containers would be more resource efficient by excluding the execution of hypervisor and guest OS, and more time efficient by avoiding booting (and shutting down) a whole OS \cite{Anderson_2015,Merkel_2014}. Nevertheless, it has been identified that the cascading layers of container images come with inherent complexity and performance penalty \cite{Banerjee_2014}. In other words, the container-based virtualization technology could also negatively impact the corresponding QoS due to its performance overhead.

Although the performance advantage of containers were investigated in several pioneer studies \cite{Walters_Chaudhary_2008,Che_Shi_2010,Xavier_Neves_2014}, the container-based virtualization solution did not gain significant popularity until the recent underlying improvements in the Linux kernel, and especially until the emergence of Docker \cite{Strauss_2013}. Starting from an open-source project in early 2013 \cite{Merkel_2014}, Docker quickly becomes the most popular container solution \cite{Pahl_2015} by significantly facilitating the management of containers. Technically, through offering the unified tool set and API, Docker relieves the complexity of utilizing the relevant kernel-level techniques including the LXC, the cgroup and a copy-on-write filesystem. To examine the performance of Docker containers, a molecular modeling simulation software \cite{Adufu_Choi_2015} and a postgreSQL database-based Joomla application \cite{Joy_2015} have been used to benchmark the Docker environment against the VM environment.


The closest work to ours is the CPU-oriented study \cite{Seo_Hwang_2014} and the IBM research report \cite{Felter_Ferreira_2015} on the performance comparison of VM and Linux containers. However, both studies are incomplete (e.g., the former was not concerned with the non-CPU features, and the latter did not finish the container's network evaluation). More importantly, our work denies the IBM report's finding that ``containers and VMs impose almost no overhead on CPU and memory usage" and also doubts about ``Docker equals or exceeds KVM performance in every case". Furthermore, in addition to the average performance overhead of virtualization technologies, we are more concerned with their overhead in performance variability. 

Note that, although there are also performance studies on deploying containers inside VMs (e.g., \cite{Dua_Raja_2015,Piraghaj_Dastjerdi_2015}), such a redundant structure might not be suitable for an ``apple-to-apple" comparison between Docker containers and VMs, and thus we do not include this virtualization scenario in this study.

\section{Performance Evaluation Implementation}
\label{sec:Implementation}
\subsection{Performance Evaluation Methodology}
\label{sec:doknowme}

Since the comparison between the container's and the VM's performance overheads is essentially based on their performance evaluation, we define our work as a performance evaluation study that belongs to the field of ECS \cite{Feitelson_2007}. Considering that ``evaluation methodology underpins all innovation in experimental computer science" \cite{Blackburn_McKinley_2008}, we employ the methodology DoKnowMe \cite{Li_OBrien_DoKnowMe} to guide evaluation implementations in this study. DoKnowMe is an abstract evaluation methodology on the analogy of ``class" in object-oriented programming. By integrating domain-specific knowledge artefacts, DoKnowMe can be customized into specific methodologies (by analogy of ``object") to facilitate evaluating different concrete computing systems. To better structure our report, we divide our DoKnowMe-driven evaluation implementation into pre-experimental activities (cf.~Section \ref{subsec:preExperimental}) and experimental results \& analyses (cf.~Section \ref{subsec:results}).

\subsection{Pre-Experimental Activities}
\label{subsec:preExperimental}
\subsubsection{Requirement Recognition}
Following DoKnowMe, the whole evaluation implementation is essentially driven by the recognized requirements. In general, the requirement recognition is to define a set of specific requirement questions both to facilitate understanding the real-world problem and to help achieve clear statements of the corresponding evaluation purpose. In this case, the basic requirement is to give a fundamental quantitative comparison between the hypervisor-based and the container-based virtualization solutions. Since we concretize these two virtualization solutions into VMWare Workstation VMs and Docker containers  respectively, such a requirement can further be specified into two questions:

\begin{enumerate}[style = standard, labelindent=0em ,labelwidth=1.3cm, labelsep*=0.5em, leftmargin =!, label=\textit{\textbf{RQ\arabic{enumi}:}}]
  \item	How much performance overhead does a standalone Docker container introduce over its base physical machine?
  \item	How much performance overhead does a standalone VM introduce over its base physical machine?
\end{enumerate}

Considering that virtualization technologies could result in service performance variation \cite{Iosup_Yigitbasi_2011}, we are also concerned with the container's and VM's potential variability overhead besides their average performance overhead: 

\begin{enumerate}[style = standard, labelindent=0em ,labelwidth=1.3cm, labelsep*=0.5em, leftmargin =!, label=\textit{\textbf{RQ\arabic{enumi}:}}]
\setcounter{enumi}{2}
  \item	How much performance variability overhead does a standalone Docker container introduce over its base physical machine during a particular period of time?
  \item	How much performance variability overhead does a standalone VM introduce over its base physical machine during a particular period of time?
\end{enumerate}

\subsubsection{Service Feature Identification}
Recall that we treat Docker containers as an alternative type of Cloud service to VMs. By using the taxonomy of Cloud services evaluation \cite{Li_OBrien_2012_taxonomy}, we examine the communication-, computation-, memory- and storage-related QoS aspects; and then we focus on the service features including communication data throughput, computation latency, memory data throughput, and storage transaction speed and data throughput.

\subsubsection{Metrics/Benchmarks Listing and Selection}
\label{subsec:metricBenchmark}
The selection of evaluation metrics usually depends on the availability of benchmarks. According to our previous experience of Cloud services evaluation, we choose relatively lightweight and popular benchmarks to try to minimize the potential benchmarking bias, as listed in Table \ref{table_benchmark}. For example, Iperf has been identified to be able to deliver more precise results by consuming less system resources. In fact, except for STREAM that is the de facto memory evaluation benchmark included in the HPC Challenge Benchmark (HPCC) suite, the other benchmarks are all Ubuntu's built-in utilities.

\begin{table}[!t]
\renewcommand{\arraystretch}{1.2}
\caption{Metrics and Benchmarks for this Evaluation Study}
\label{table_benchmark}
\centering

\begin{tabular}{|l|c|c|c|}
\hline
\textbf{Physical Property} & \textbf{Capacity Metric} & \textbf{Benchmark} & \textbf{Version}\\
\hline
Communication & Data Throughput & Iperf & 2.0.5 \\
\hline
Computation & (Latency) Score & HardInfo & 0.5.1\\
\hline
Memory & Data Throughput & STREAM & 5.10\\
\hline
Storage & Transaction Speed & Bonnie++ & 1.97.1\\
\hline
Storage & Data Throughput & Bonnie++ & 1.97.1\\
\hline
\end{tabular}
\end{table}

In particular, although Bonnie++ only measures the amount of data processed per second, the disk I/O transactions are on a byte-by-byte basis when accessing small size of data. Therefore, we consider to measure storage transaction speed when operating byte-size data and measure storage data throughput when operating block-size data. As for the property computation, considering the diversity in CPU jobs (e.g., integer and floating-point calculations), we employ HardInfo that includes six micro-benchmarks to generate performance scores. 

When it comes to the performance overhead, we use the business domain's \textit{Overhead Ratio}\footnote{\url{http://www.investopedia.com/terms/o/overhead-ratio.asp}} as an analogy to its measurement. In detail, we treat the performance loss compared to a baseline as the expense, while imagining the baseline performance to be the overall income, as defined in Equation (\ref{eq:overhead}). 

\begin{footnotesize}
\begin{equation}
\label{eq:overhead}
O_p=\frac{\left|P_m-P_b\right|}{P_b} \times 100\%
\end{equation}
\end{footnotesize}%
where $O_p$ refers to the performance overhead; $P_m$ denotes the benchmarking result as a measurement of a service feature; $P_b$ indicates the baseline performance of the service feature; and then $\left|P_m-P_b\right|$ represents the corresponding performance loss. Note that the physical machine's performance is used as the baseline in our study. Moreover, considering possible observational errors, we allow a margin of error for the confidence level as high as 99\% with regarding to the benchmarking results. In other words, we will ignore the difference between the measured performance and its baseline if the calculated performance overhead is less than 1\% (i.e.~if $O_p<1\%$, then $P_m=P_b$). 

\subsubsection{Experimental Factor Listing and Selection}
The identification of experimental factors plays a prerequisite role in the following experimental design. More importantly, specifying the relevant factors would be necessary for improving the repeatability of experimental implementations. By referring to the experimental factor framework of Cloud services evaluation \cite{Li_OBrien_2012_factor}, we choose the resource- and workload-related factors as follows. In particular, considering the possible performance overhead compensation from powerful computing resources, we try to stress the experimental condition by employing a ``just-enough" testbed.

\textbf{The resource-related factors:}

\begin{itemize}
	\item \textit{Resource Type:} Given the evaluation requirement, we have essentially considered three types of resources to support the imaginary Cloud service, namely physical machine, container and VM.
	\item \textit{Communication Scope:} We test the communication between our local machine and an Amazon EC2 t2.micro instance. The local machine is located in our broadband lab at Lund University, and the EC2 instance is from Amazon's available zone ap-southeast-1a within the region Asia Pacific (Singapore). 
	\item \textit{Communication Ethernet Index:} Our local side uses a Gigabit connection to the Internet, while the EC2 instance at remote side has the ``Low to Moderate" networking performance defined by Amazon.
	\item \textit{CPU Index:} Recall that we have employed ``just-enough" computing resource. The physical machine's CPU model is chosen to be Intel Core\texttrademark 2 Duo Processor T7500. The processor has two cores with the 64-bit architecture, and its base frequency is 2.2 GHz. We allocate both CPU cores to the standalone VM upon the physical machine. 
	\item \textit{Memory Size:} The physical machine is equipped with a 3GB DDR2 SDRAM. When running the VMWare Workstation Pro without launching any VM, ``\texttt{watch -n 5 free -m}" shows a memory usage of 817MB while leaving 2183MB free in the physical machine. Therefore, we set the memory size to 2GB for the VM to avoid (at least to minimize) the possible memory swapping.
	\item \textit{Storage Size:} There are 120GB of hard disk in the physical machine. Considering the space usage by the host operating system, we allocate 100GB to the VM.
	\item \textit{Operating System:} Since Docker requires a 64-bit installation and Linux kernels older than 3.10 do not support all the features for running Docker containers, we choose the latest 64-bit Ubuntu 15.10 as the operating system for both the physical machine and the VM. In addition, according to the discussions about base images in the Docker community \cite{StackOverflow_2013,Reddit_2015}, we intentionally set an OS base image (by specifying \texttt{FROM ubuntu:15.10} in the Dockerfile) for all the Docker containers in our experiments. Note that a container's OS base image is only a file system representation, while not acting as a guest OS.
\end{itemize}

\textbf{The workload-related factors:}

\begin{itemize}
	\item \textit{Duration:} For each evaluation experiment, we decided 
to take a whole-day observation plus one-hour warming up (i.e.~25 hours).
	\item \textit{Workload Size:} The experimental workloads are predefined by the selected benchmarks. For example, the micro-benchmark CPU Fibonacci generates workload by calculating the $42nd$ Fibonacci number. In particular, the benchmark Bonnie++ distinguishes between reading/writing byte-size and block-size data.
\end{itemize}

\subsubsection{Experimental Design}
It is clear that the identified factors are all with single value except for the \textit{Resource Type}. Therefore, a straightforward design is to run the individual benchmarks on each of the three types of resources independently for a whole day plus one hour.

Furthermore, following the conceptual model of IaaS performance evaluation \cite{Li_OBrien_2014_blueprint}, we record the experimental design into a blueprint both to facilitate our experimental implementations and to help other evaluators replicate/repeat our study. Due to the space limit, we share the experimental blueprint online as a supplementary document.\footnote{The experimental blueprint is shared online at \url{https://drive.google.com/file/d/0B9KzcoAAmi43WTFuTXBsZ0NRd1U/}}

\subsection{Experimental Results and Analyses}
\label{subsec:results}
\subsubsection{Communication Evaluation Result and Analysis}

For the purpose of ``apple-to-apple" comparison, we force both the container and the VM to employ Network Address Translation (NAT) to establish outgoing connections. Since they require port binding/forwarding to accept incoming connections, we only test the outgoing communication performance to reduce the possibility of configurational noise, by setting the remote EC2 instance to Iperf server and using the local machine, container and VM all as Iperf clients.

The benchmarking results of repeating \texttt{iperf -c XXX.XXX.XXX.XXX -t 15} (with a one-minute interval between every two consecutive trials) are listed in Table \ref{table_iperf}. The \texttt{XXX.XXX.XXX.XXX} denotes the external IP address of the EC2 instance used in our experiments. Note that, unlike the other performance features, the communication data throughput delivers periodical and significant fluctuations, which might be a result from the network resource competition at both our local side and the EC2 side during working hours.  Therefore, we particularly focus on the longest period of relatively stable data out of the whole-day observation, and thus the results here are for rough reference only.

\begin{table}[!t]\footnotesize
\renewcommand{\arraystretch}{1.2}
\caption{Communication Benchmarking Results using Iperf}
\label{table_iperf}
\centering
\begin{tabular}{|l|c|c|c|c|}
\hline
\textbf{Resource Type} & \textbf{Average}& \textbf{Standard Deviation}\\
\hline
Physical machine& 29.066 Mbits/sec& 1.282 Mbits/sec\\
\hline
Container& 28.484 Mbits/sec&	1.978 Mbits/sec\\
\hline
Virtual machine & 12.843 Mbits/sec&	2.979 Mbits/sec\\
\hline
\end{tabular}
\end{table}

\begin{figure}[!t]
  \centering
  \includegraphics[width=7.4cm]{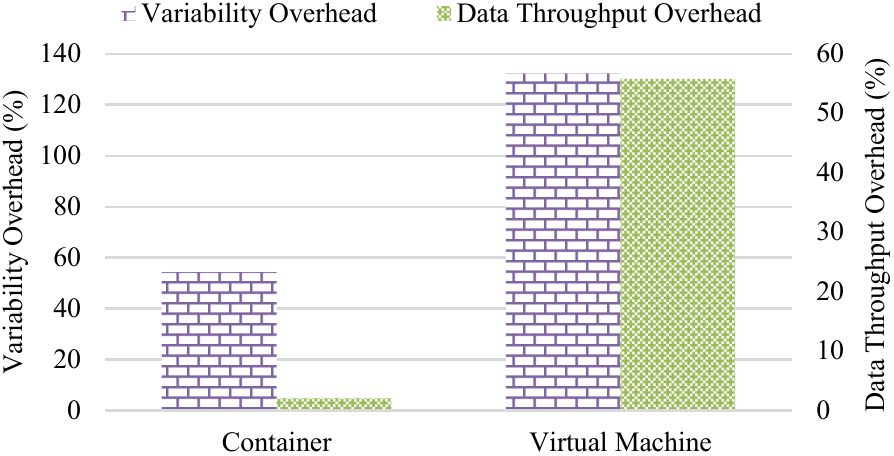}
  \caption{Communication data throughput and its variability overhead of a standalone Docker container vs.~VM (using the benchmark Iperf).}
  \label{fig:iperfOverhead} 
\end{figure}

Given the extra cost of using the NAT network to send and receive packets, there would be unavoidable performance penalties for both the container and the VM. Using Equation (\ref{eq:overhead}), we calculate their communication performance overheads, as illustrated in Figure \ref{fig:iperfOverhead}.

A clear trend is that, compared to the VM, the container loses less communication performance, with only $2\%$ data throughput overhead and around $54\%$ variability overhead. However, it is surprising to see a more than $55\%$ data throughput overhead for the VM. Although we have double checked the relevant configuration parameters and redone several rounds of experiments to confirm this phenomenon, we still doubt about the hypervisor-related reason behind such a big performance loss. We particularly highlight this observation to inspire further investigations.

\subsubsection{Computation Evaluation Result and Analysis}
Recall that HardInfo's six micro benchmarks deliver both ``higher=better" and ``lower=better" CPU scores. To facilitate experimental analysis, we use the two equations below to standardize the ``higher=better" and ``lower=better" benchmarking results respectively. 

\begin{footnotesize}
\begin{equation}
\label{eq:higherbetter}
\textit{HB}_i=\frac{Benchmarking_i}{\max(Benchmarking_{1,2, ..., n})}
\end{equation}
\end{footnotesize}
\begin{footnotesize}
\begin{equation}
\label{eq:lowerbetter}
\textit{LB}_i=\cfrac{\cfrac{1}{Benchmarking_i}}{\max(\cfrac{1}{Benchmarking_{1,2, ..., n}})}
\end{equation}
\end{footnotesize}%
where $\textit{HB}_i$ further scores the service resource type $i$ by standardizing the ``higher=better" benchmarking result $Benchmarking_i$; and similarly, $\textit{LB}_i$ represents the standardized ``lower=better" CPU score of the service resource type $i$. Note that Equation (\ref{eq:lowerbetter}) essentially offers the ``lower=better" benchmarking results a ``higher=better" representation through reciprocal standardization. 

\begin{figure}[!t]
  \centering
  \includegraphics[width= 7.4cm]{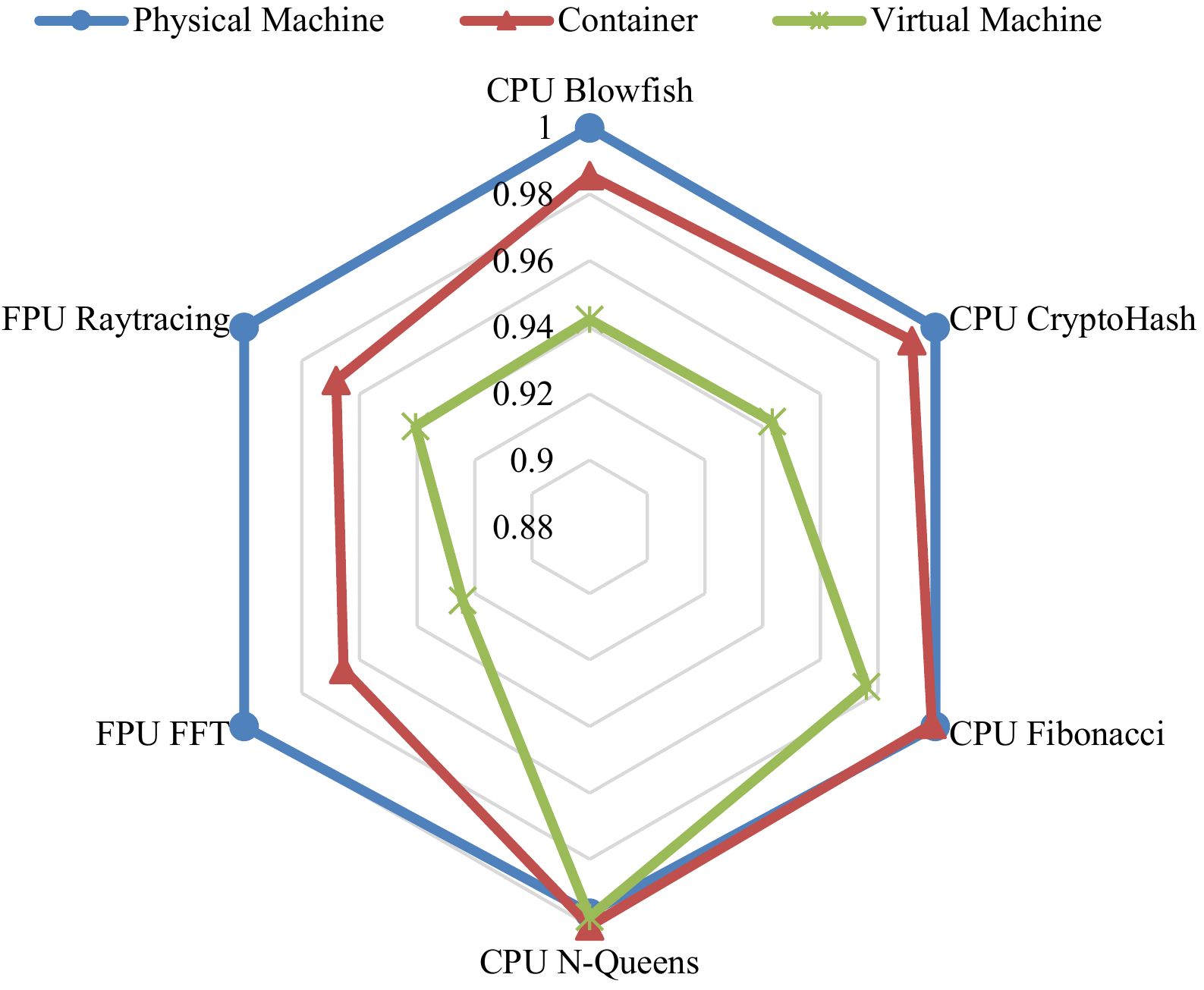}
  \caption{Computation benchmarking results by using HardInfo.}
  \label{fig:cpufeature} 
\end{figure}

\begin{figure}[!t]
  \centering
  \includegraphics[width= 7.4cm]{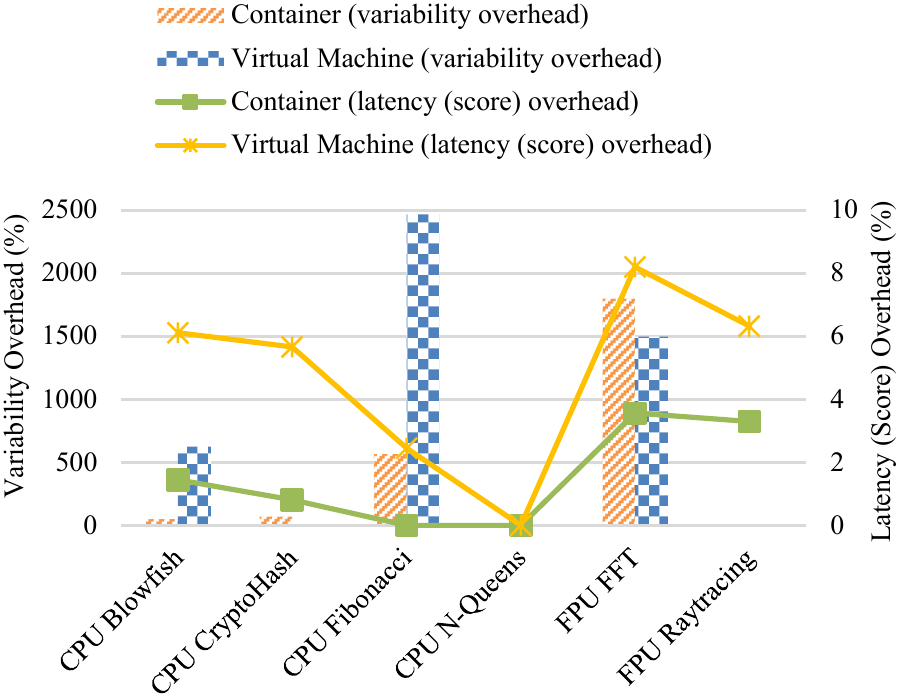}
  \caption{Computation latency (score) and its variability overhead of a standalone Docker container vs.~VM (using the tool kit HardInfo).}
  \label{fig:cpuoverhead} 
\end{figure}

Thus, we can use a radar plot to help intuitively contrast the performance of the three resource types, as demonstrated in Figure \ref{fig:cpufeature}. For example, the different polygon sizes clearly indicate that the container generally computes faster than the VM, although the performance differences are on a case-by-case basis with respect to different CPU job types.

Nevertheless, our experimental results do not display any general trend in variability of those resources' computation scores. As can be seen from the calculated performance overheads (cf.~Figure \ref{fig:cpuoverhead}), the VM does not even show worse variability than the physical machine when running CPU CryptoHash, CPU N-Queens and FPU Raytracing. On the contrary, there is an almost $2500\%$ variability overhead for the VM when calculating the $42nd$ Fibonacci number. In particular, the virtualization technologies seem to be sensitive to the Fourier transform jobs (the benchmark FPU FFT), because the computation latency overhead and the variability overhead are relatively high for both the container and the VM.

\subsubsection{Memory Evaluation Result and Analysis}
\begin{figure}[!t]
  \centering
  \includegraphics[width= 7.4cm]{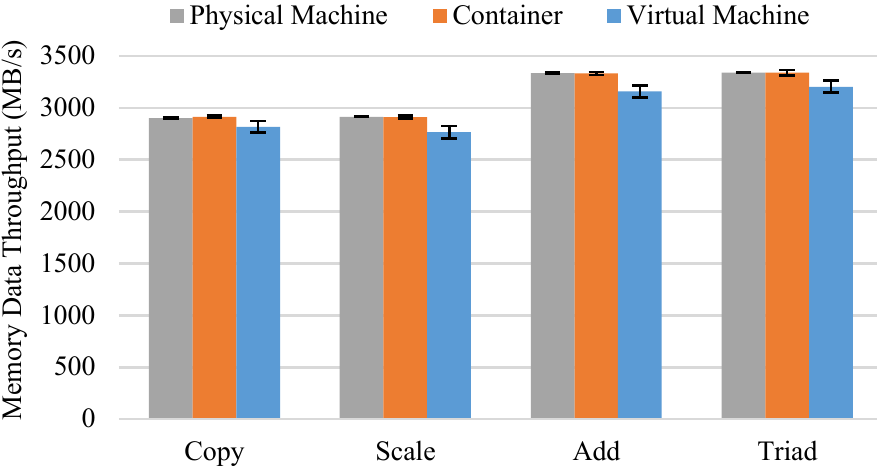}
  \caption{Memory benchmarking results by using STREAM. Error bars indicate the standard deviations of the corresponding memory data throughput.}
  \label{fig:memory} 
\end{figure}

STREAM measures sustainable memory data throughput by conducting four typical vector operations, namely Copy, Scale, Add and Triad. We directly visualize the benchmarking results into Figure \ref{fig:memory} to facilitate our observation. As the first impression, it seems that the VM has a bit poorer memory data throughput, and there is little difference between the physical machine and the Docker container in the context of running STREAM.

\begin{figure}[!t]
  \centering
  \includegraphics[width= 7.4cm]{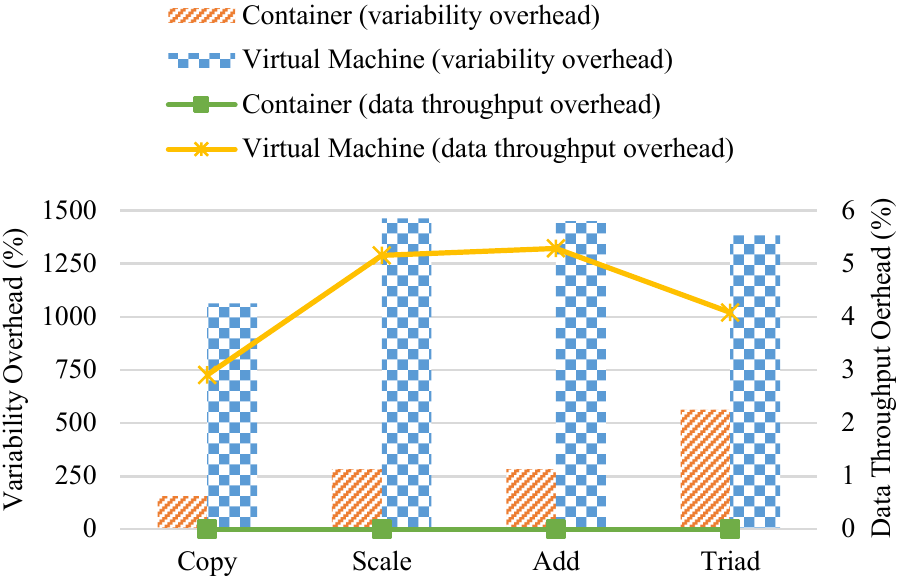}
  \caption{Memory data throughput and its variability overhead of a standalone Docker container vs.~VM (using the benchmark STREAM).}
  \label{fig:memoryOverhead} 
\end{figure}

By calculating the performance overhead in terms of memory data throughput and its variability, we are able to see the significant difference among these three types of resources, as illustrated in Figure \ref{fig:memoryOverhead}. Take the operation Triad as an example, although the container performs as well as the physical machine on average, the variability overhead of the container is more than $500\%$; similarly, although the VM's Triad data throughput overhead is around $4\%$ only, its variability overhead is almost $1400\%$. In other words, the memory performance loss incurred by both virtualization techniques is mainly embodied with the increase in the performance variability.

In addition, it is also worth notable that the container's average Copy data throughput is even slightly higher than the physical machine (i.e.~$2914.023$MB/s vs.~$2902.685$MB/s) in our experiments. Recall that we have considered a $1\%$ margin of error. Since those two values are close to each other within this error margin, here we ignore such an irregular phenomenon as an observational error.

\begin{figure*}[!t]
  \centering
  \subfloat[Physical machine writes bytes.]{
    \label{fig:subfigMachineWriteChar} 
    \includegraphics[width=4cm]{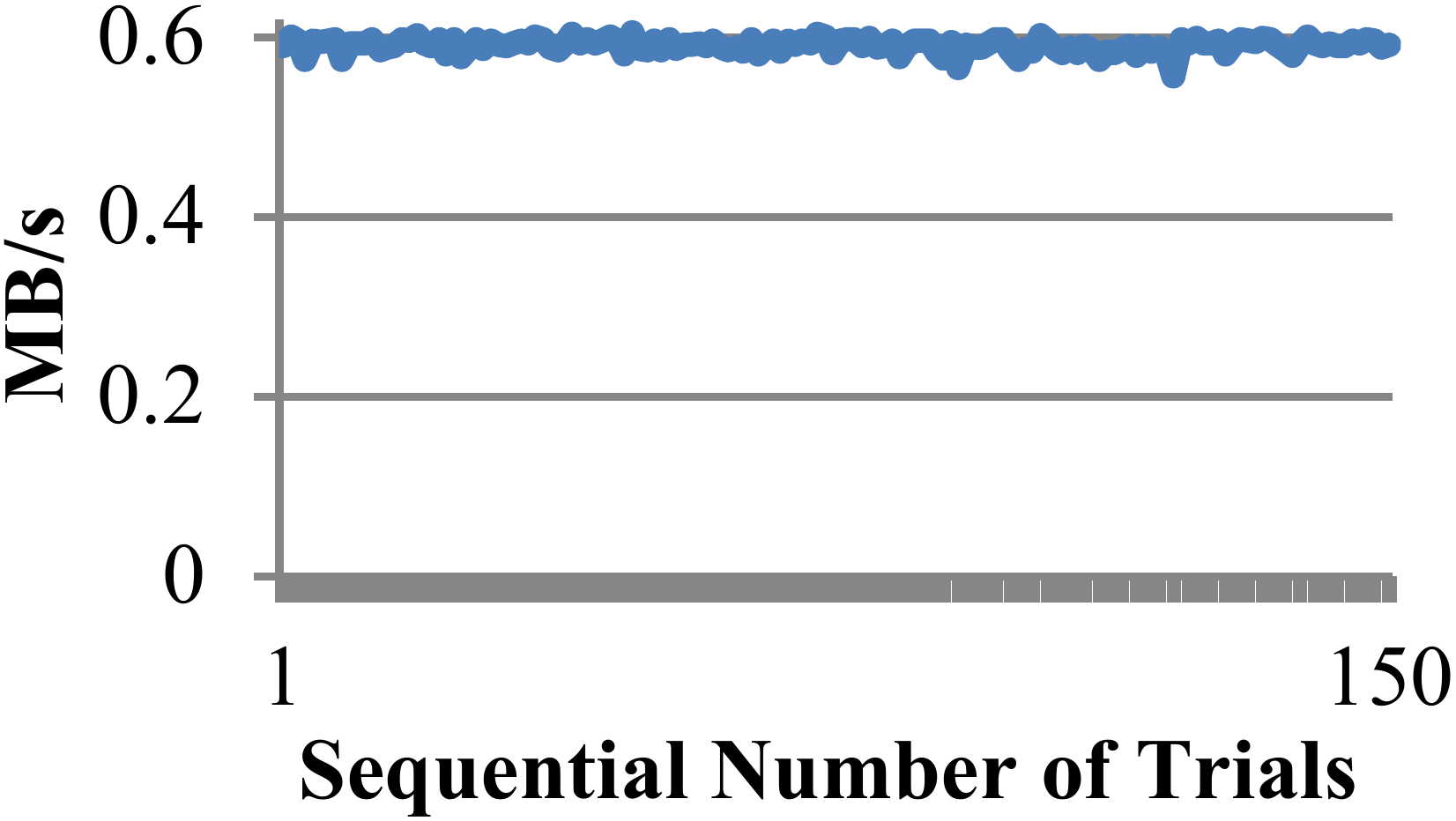}}
  \subfloat[Physical machine writes blocks.]{
    \label{fig:subfigMachineWriteBlock} 
    \includegraphics[width=4cm]{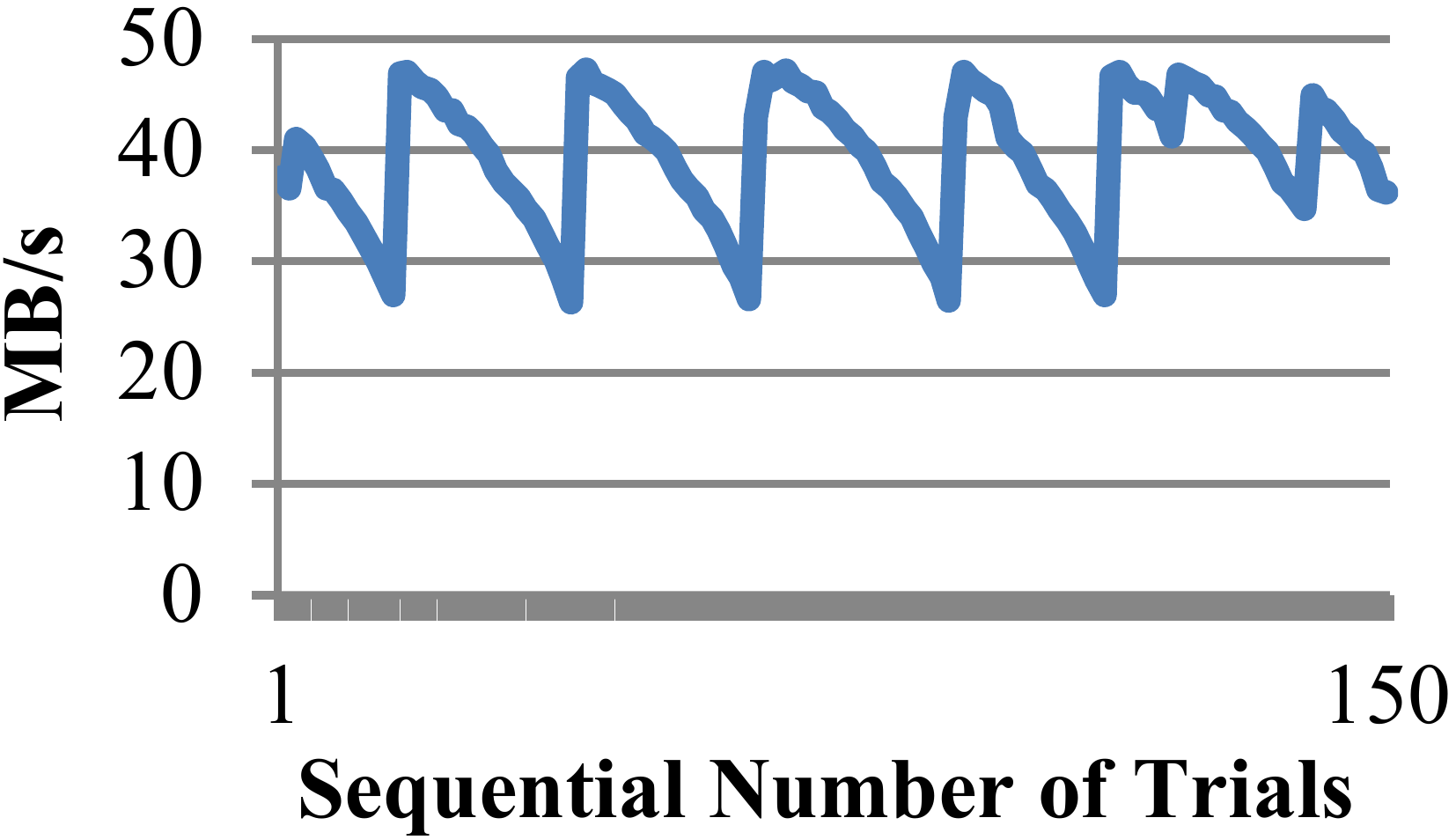}}
  \subfloat[Physical machine reads bytes.]{
    \label{fig:subfigMachineReadChar} 
    \includegraphics[width=4cm]{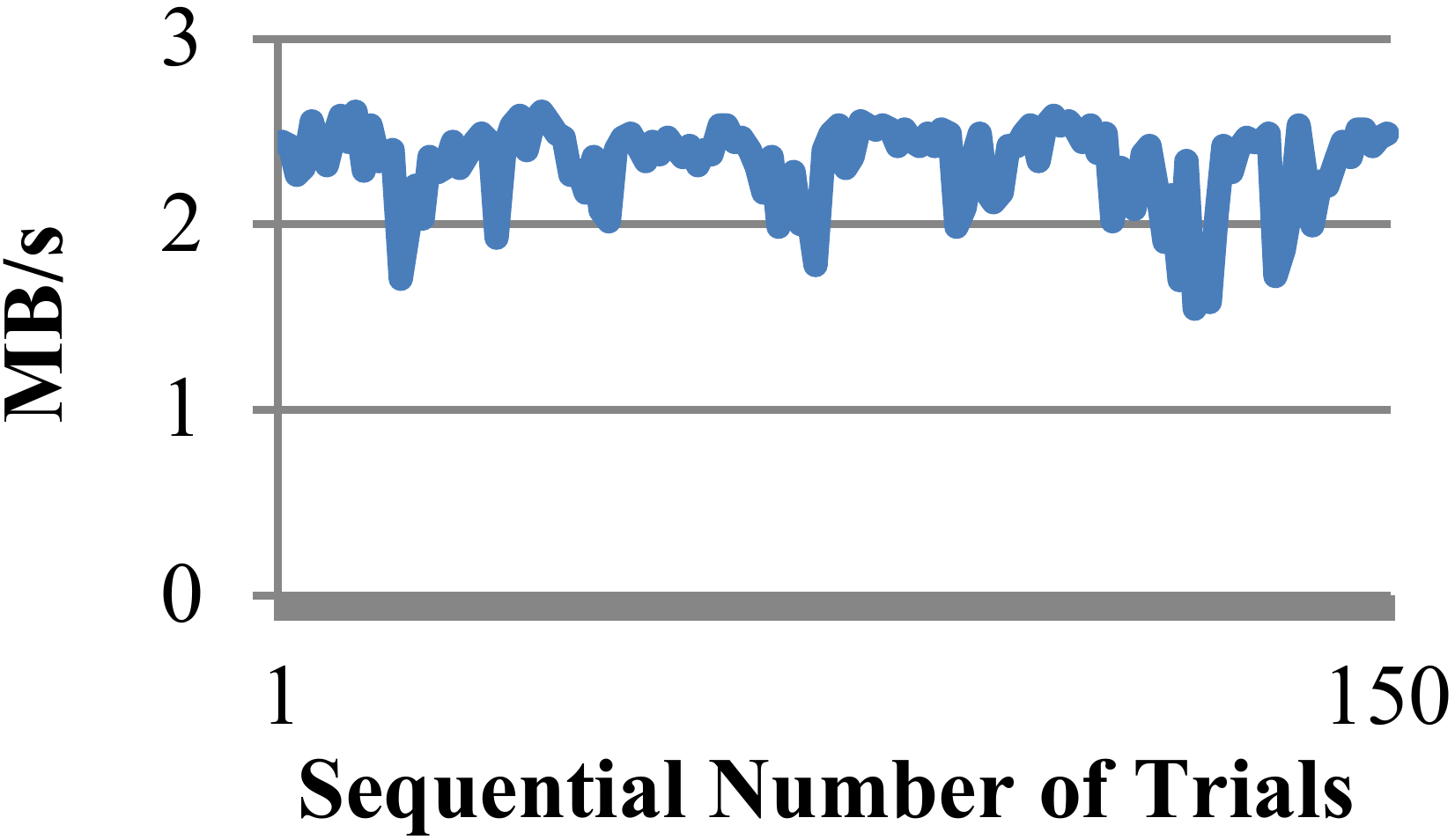}}
  \subfloat[Physical machine reads blocks.]{
    \label{fig:subfigMachineReadBlock} 
    \includegraphics[width=4cm]{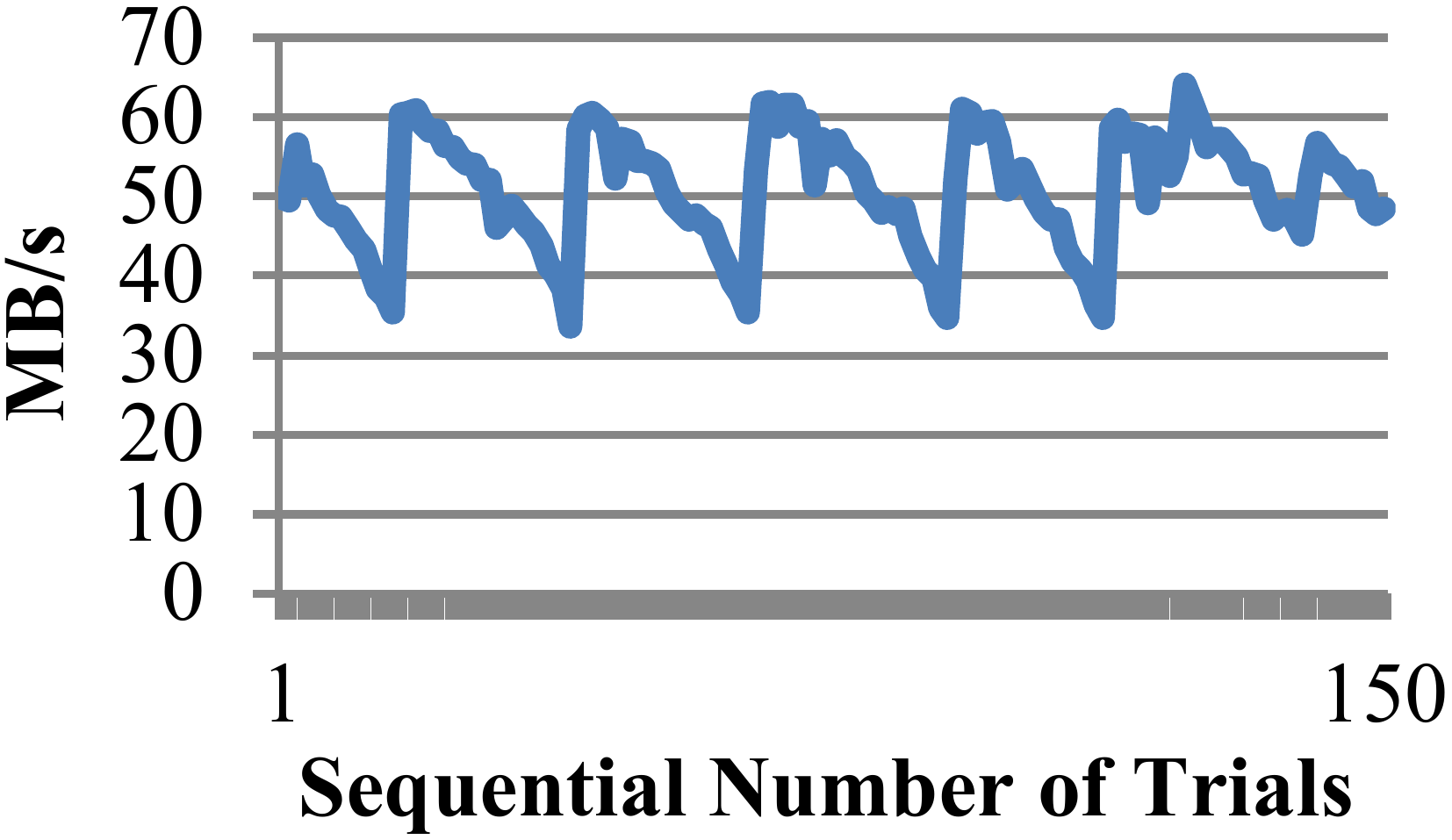}}

  \subfloat[Container writes bytes.]{
    \label{fig:subfigContainerWriteChar} 
    \includegraphics[width=4cm]{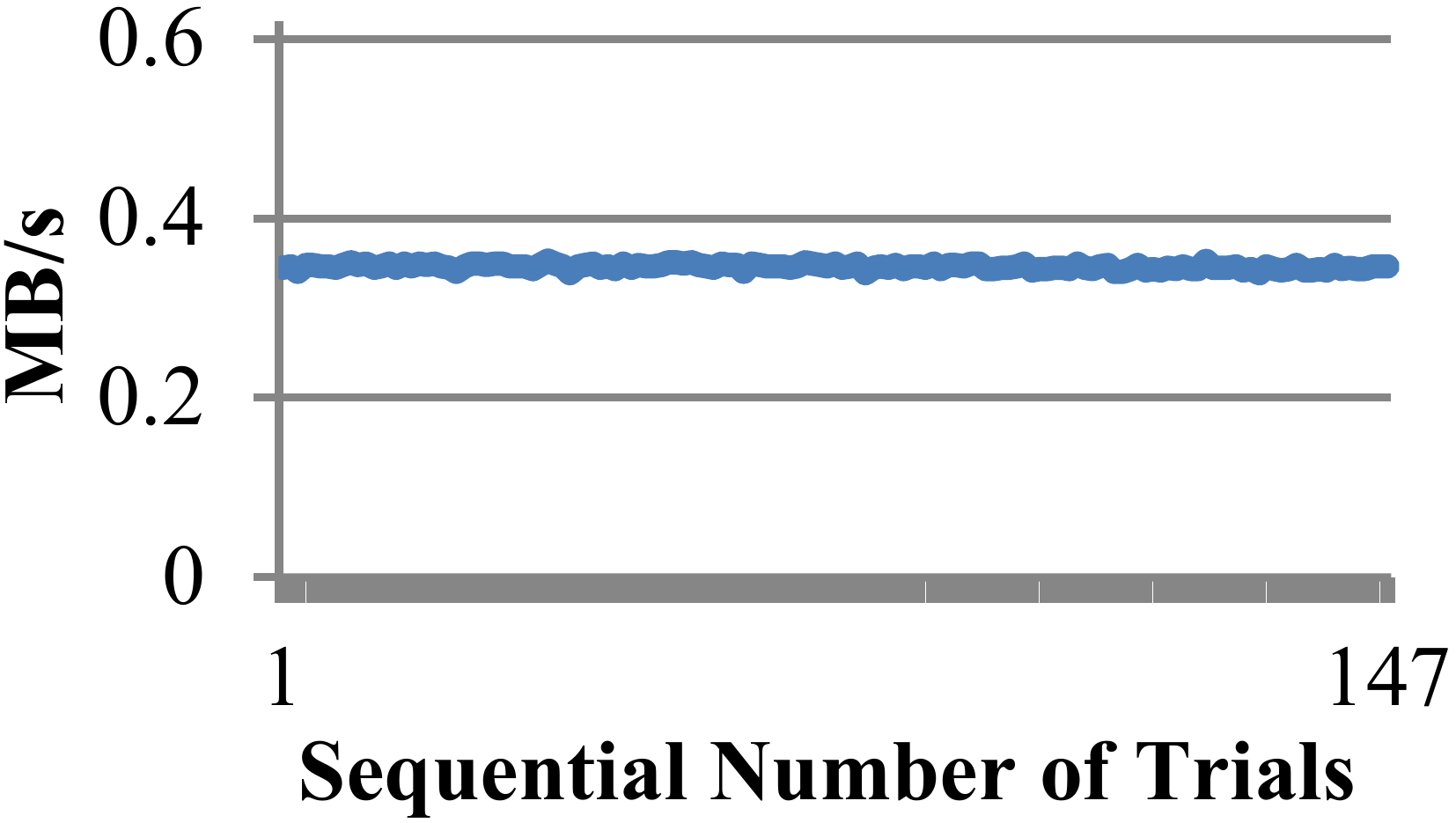}}
  \subfloat[Container writes blocks.]{
    \label{fig:subfigContainerWriteBlock} 
    \includegraphics[width=4cm]{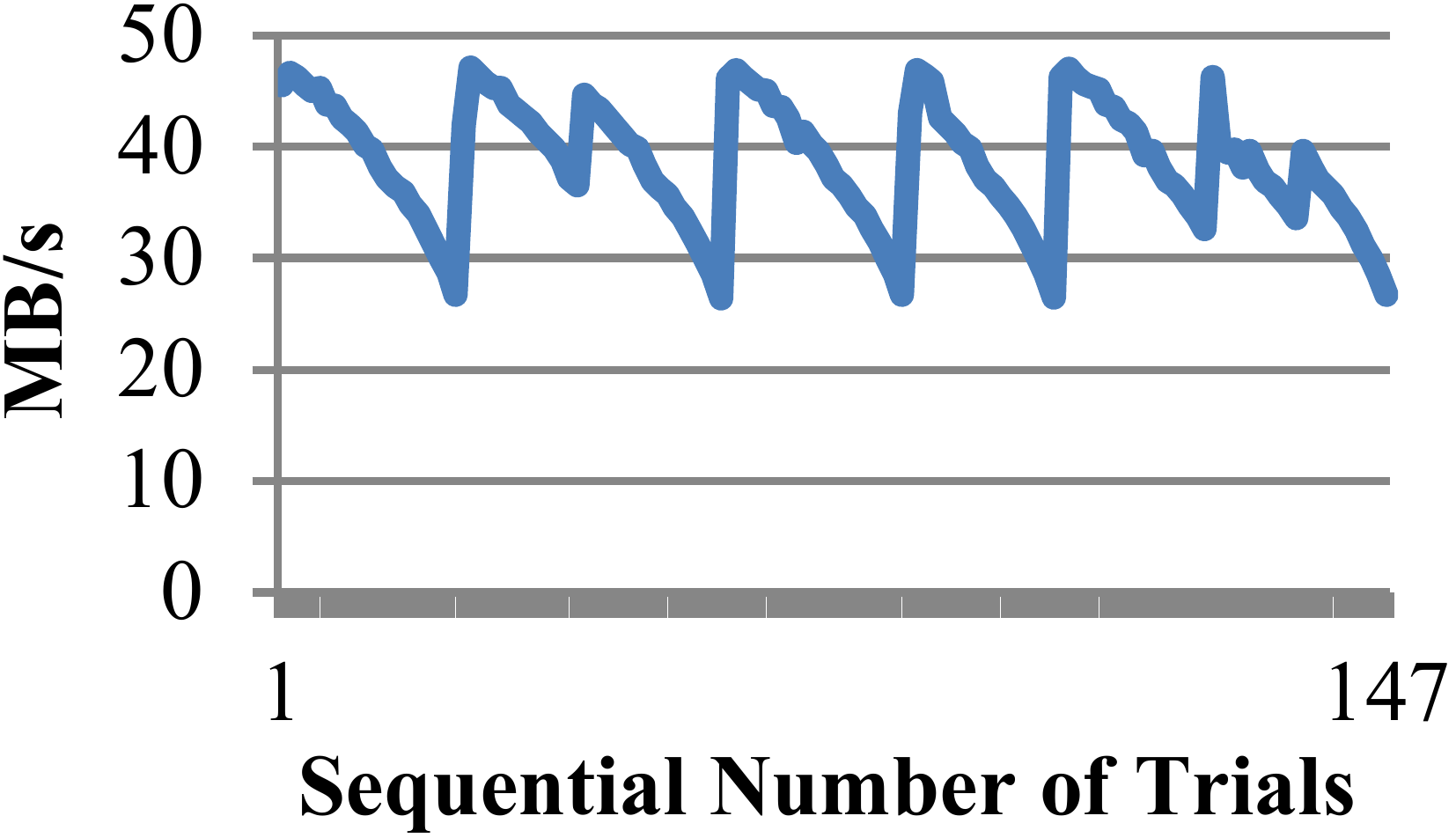}}
  \subfloat[Container reads bytes.]{
    \label{fig:subfigContainerReadChar} 
    \includegraphics[width=4cm]{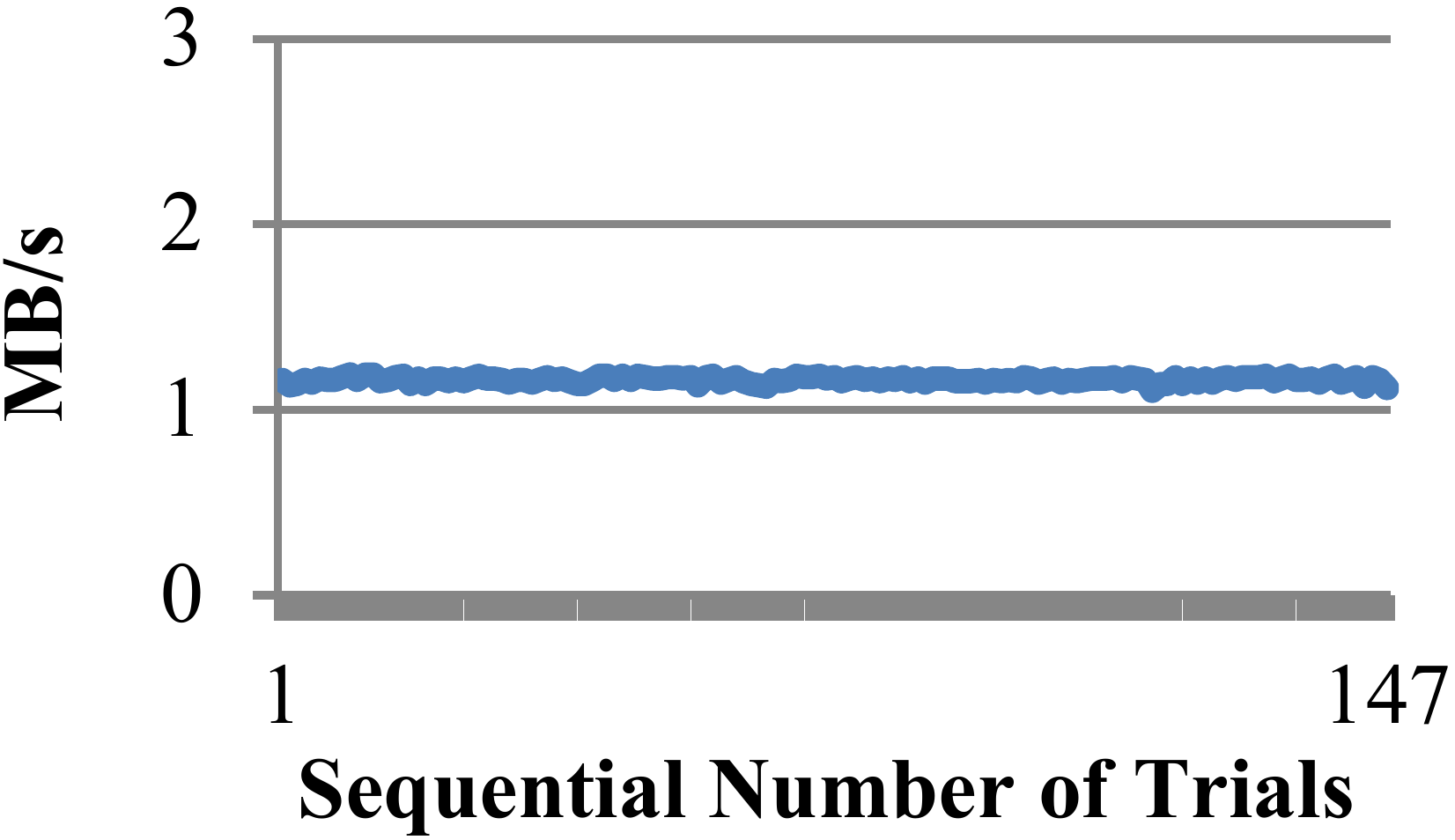}}
  \subfloat[Container reads blocks.]{
    \label{fig:subfigContainerReadBlock1} 
    \includegraphics[width=4cm]{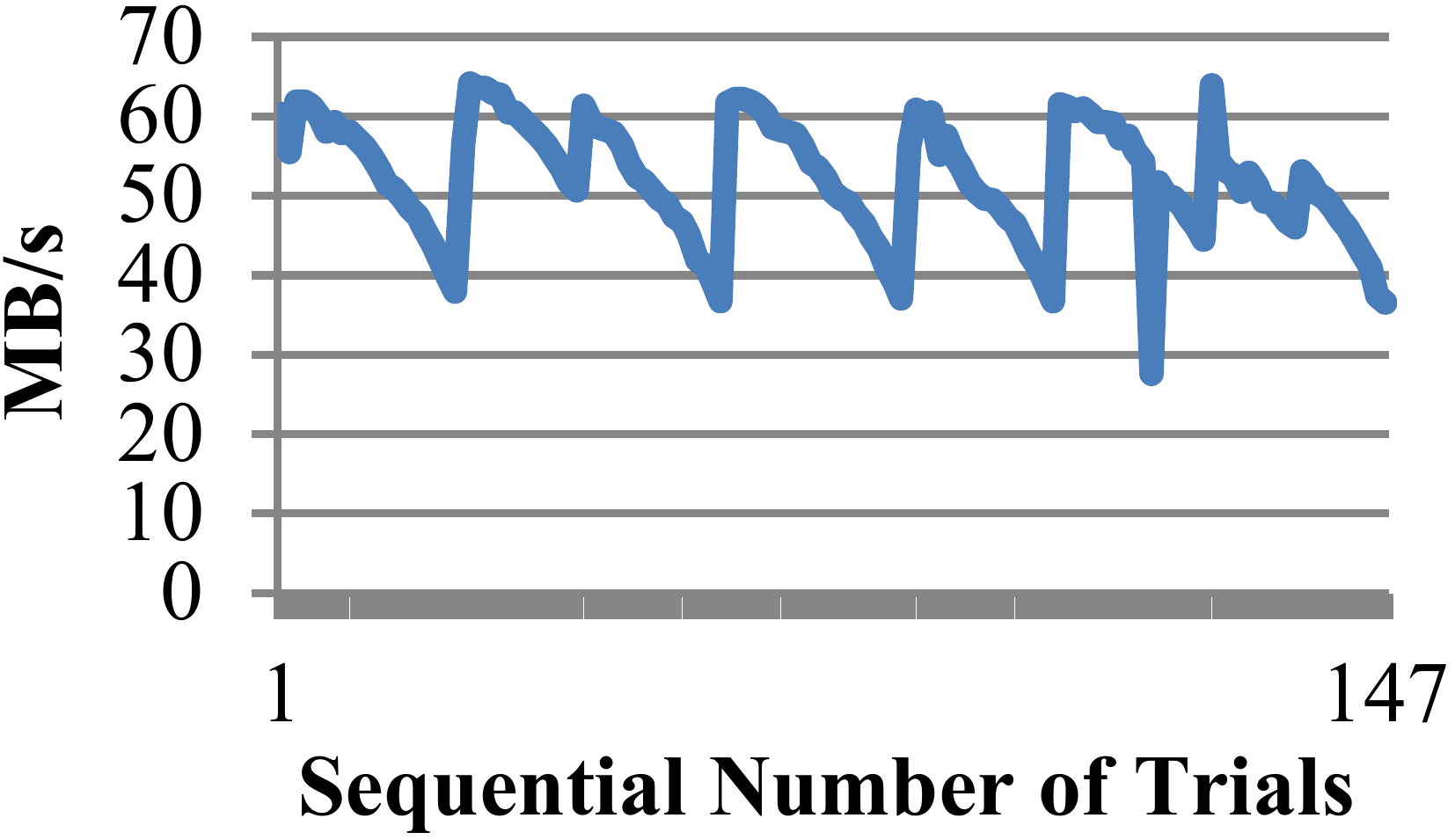}}

  \subfloat[Virtual machine writes bytes.]{
    \label{fig:subfigVmWriteChar} 
    \includegraphics[width=4cm]{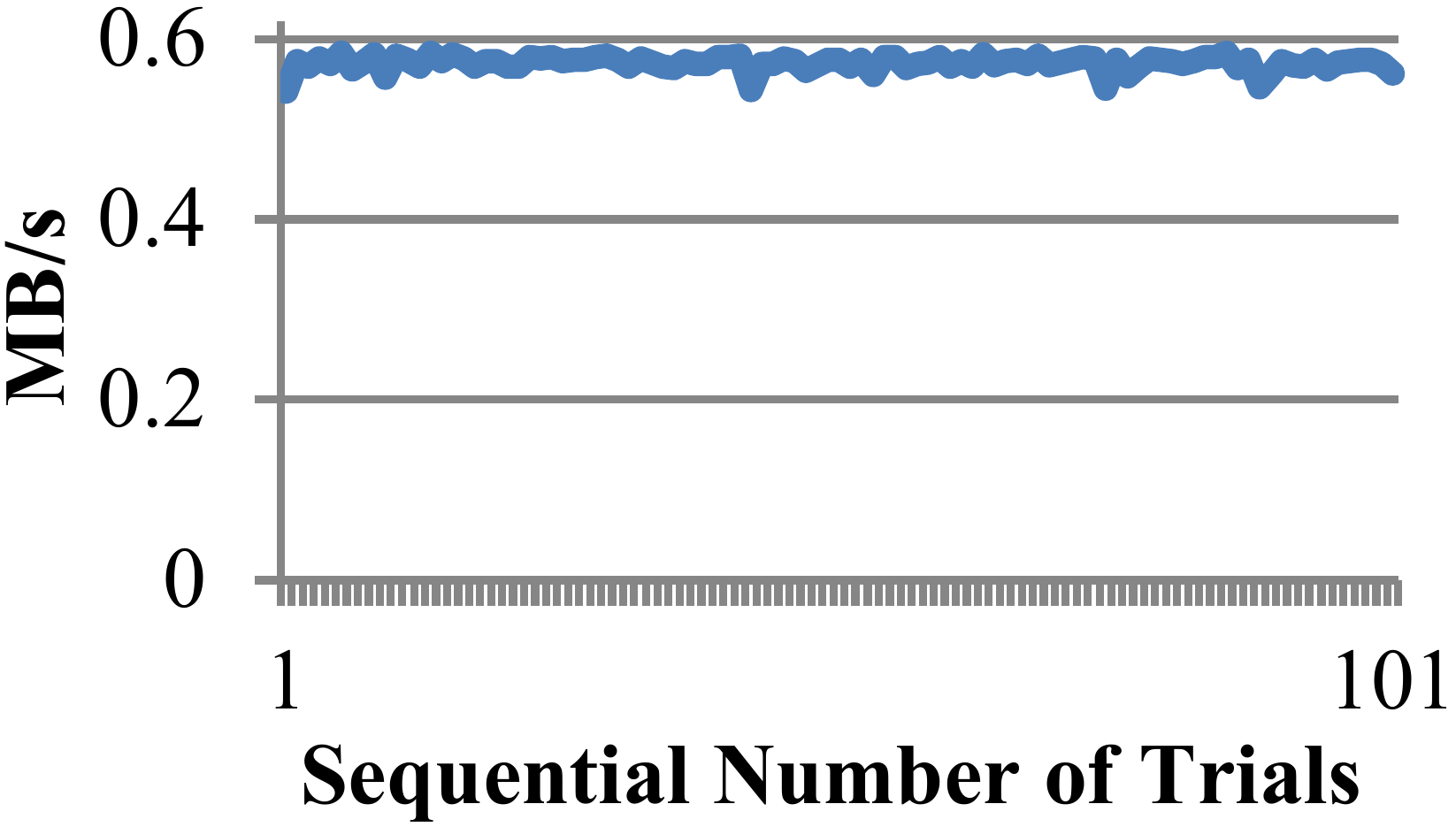}}
  \subfloat[Virtual machine writes blocks.]{
    \label{fig:subfigVmWriteBlock} 
    \includegraphics[width=4cm]{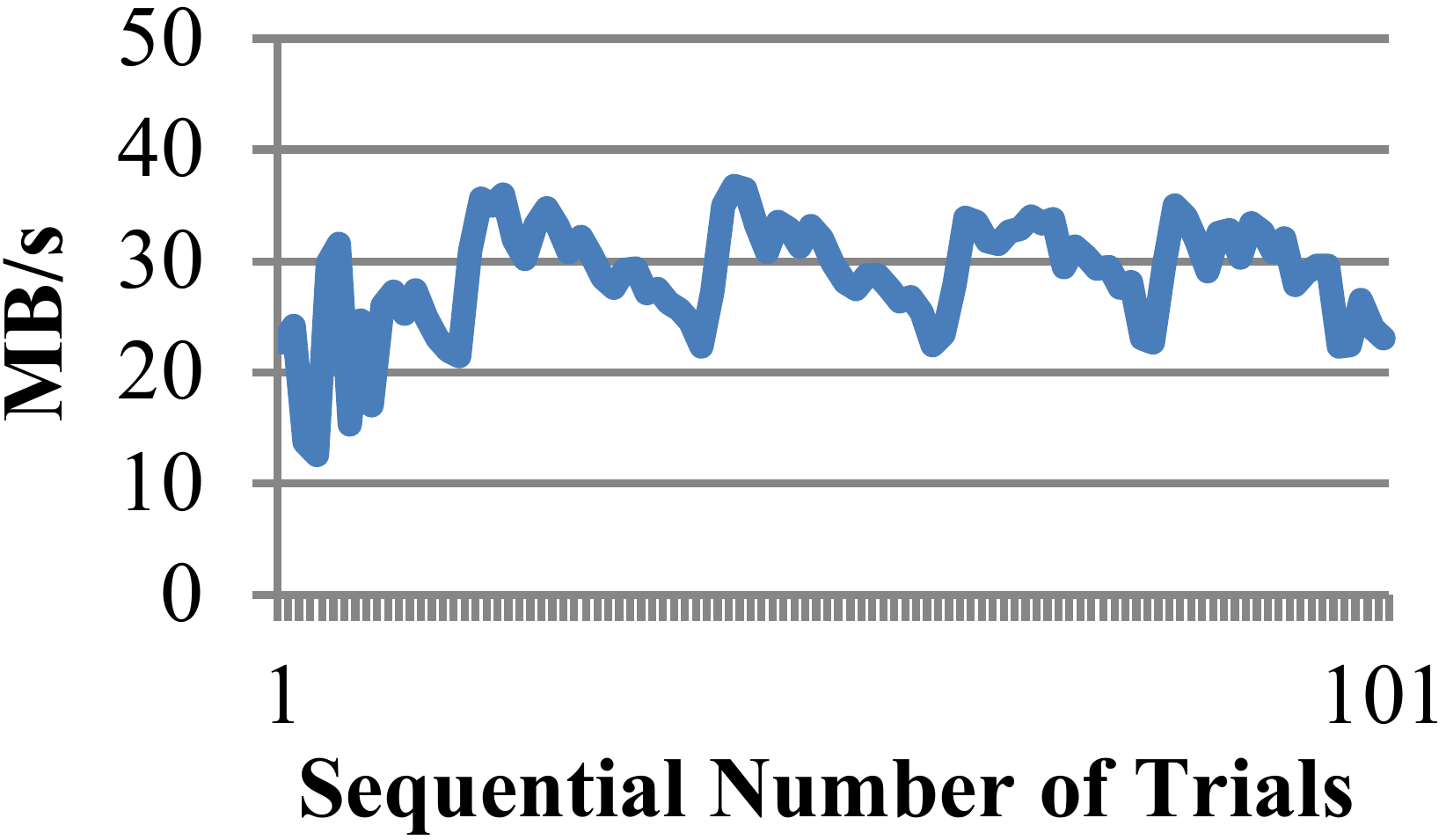}}
  \subfloat[Virtual machine reads bytes.]{
    \label{fig:subfigVmReadChar} 
    \includegraphics[width=4cm]{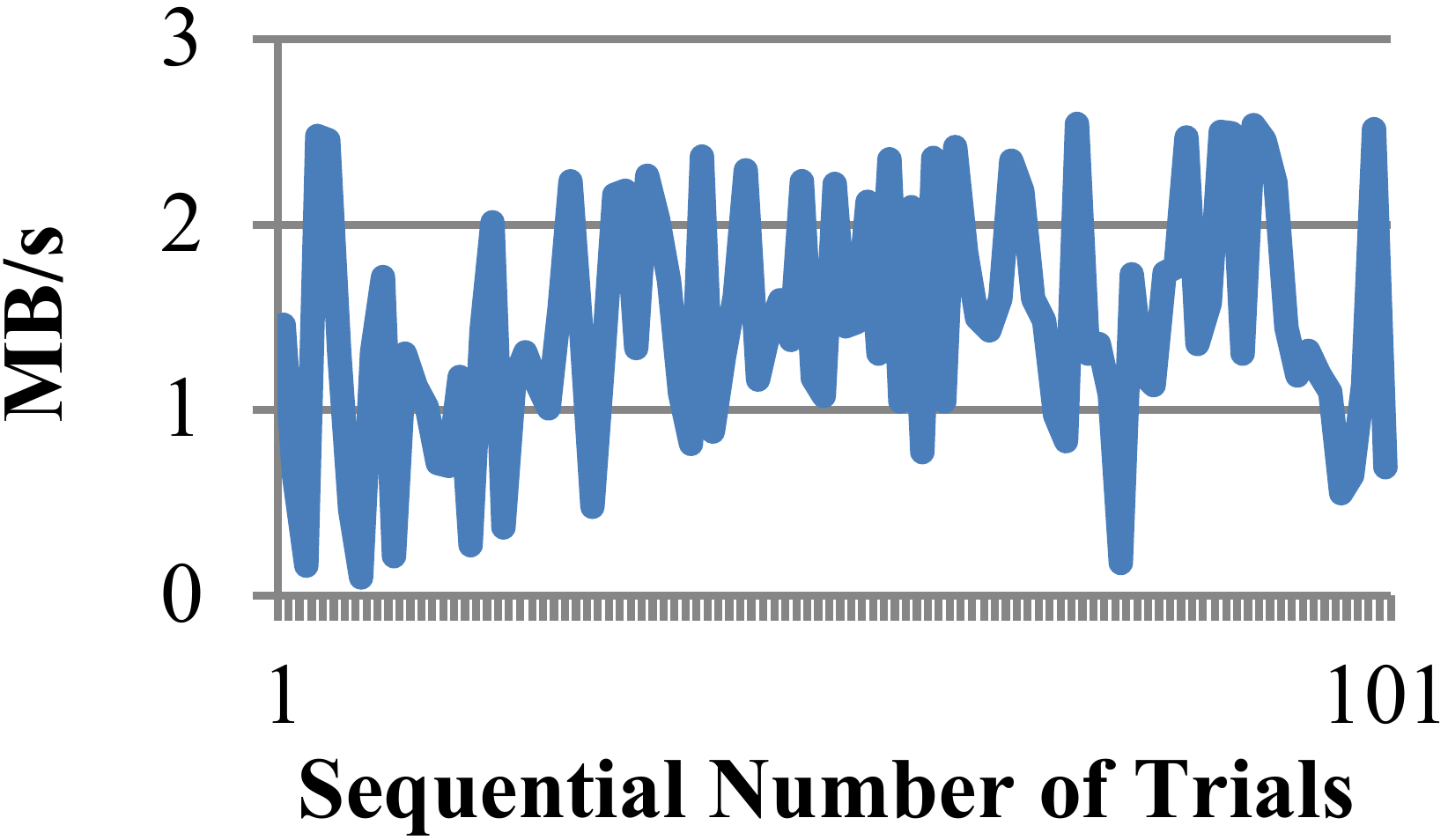}}
  \subfloat[Virtual machine reads blocks.]{
    \label{fig:subfigVmReadBlock1} 
    \includegraphics[width=4cm]{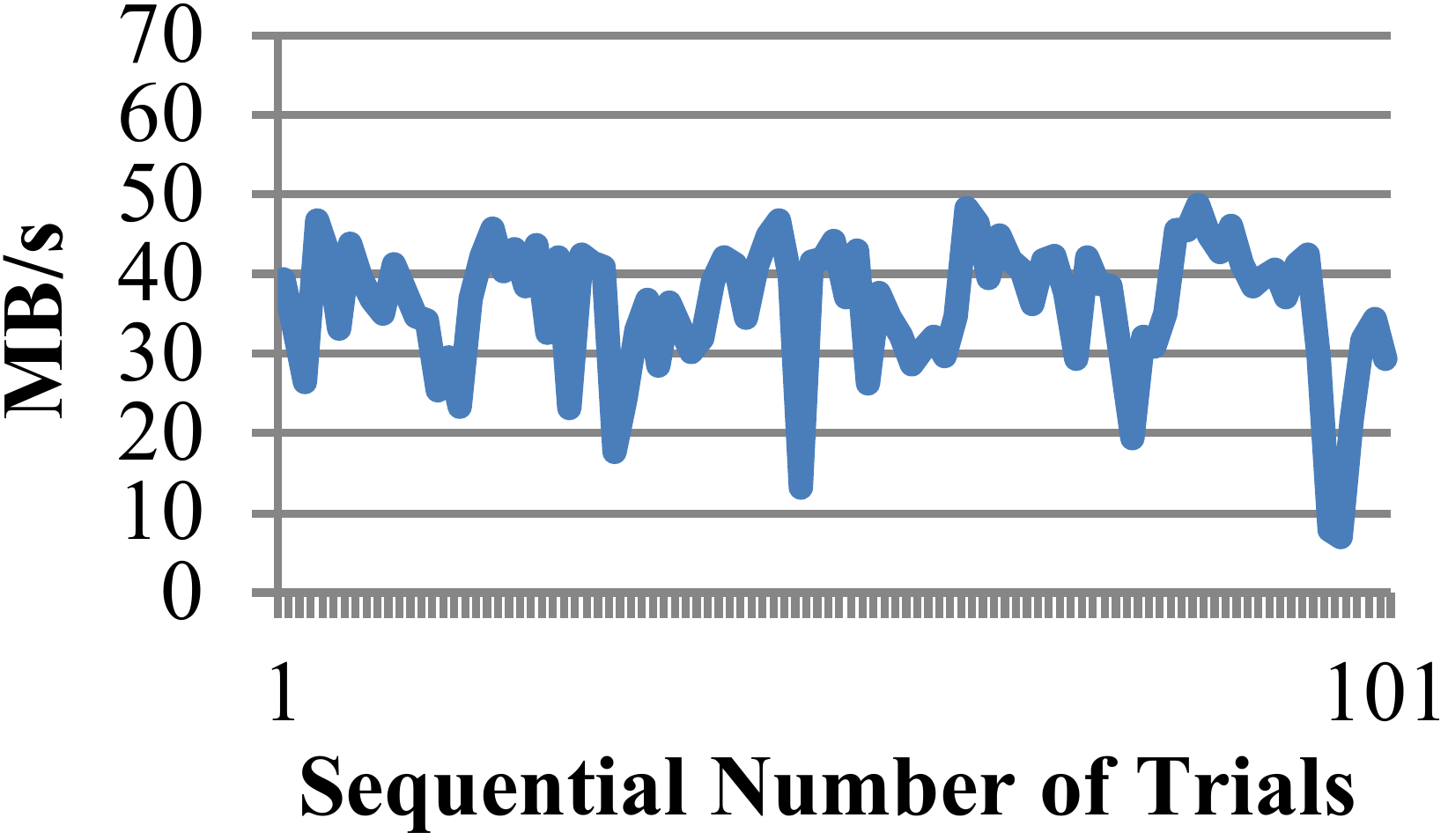}}

  \caption{Storage benchmarking results by using Bonnie++ during 24 hours. The maximum x-axis scale indicates the iteration number of the Bonnie++ test (i.e.~the physical machine, the container and the VM run 150, 147 and 101 tests respectively).}
  \label{fig:bonnieRun} 
\end{figure*}

\subsubsection{Storage Evaluation Result and Analysis}
For the test of disk reading and writing, Bonnie++ creates a dataset twice the size of the involved RAM memory. Since the VM is allocated 2GB of RAM, we also restrict the memory usage to 2GB for Bonnie++ on both the physical machine and the container, by running ``\texttt{sudo bonnie++ -r 2048 -n 128 -d / -u root}". Correspondingly, the benchmarking trials are conducted with 4GB of random data on the disk. When Bonnie++ is running, it carries out various storage operations ranging from data reading/writing to file creating/deleting. Here we only focus on the performance of reading/writing byte- and block-size data. 

To help highlight several different observations, we plot the trajectory of the experimental results along the trial sequence during the whole day, as shown in Figure \ref{fig:bonnieRun}. The first surprising observation is that, all the three resource types have regular patterns of performance jitter in block writing, rewriting and reading. Due to the space limit, we do not report their block rewriting performance in this paper. By exploring the hardware information, we identified the hard disk drive (HDD) model to be ATA Hitachi HTS54161, and its specification describes ``It stores 512 bytes per sector and uses four data heads to read the data from two platters, rotating at 5,400 revolutions per minute". As we know, the hard disk surface is divided into a set of concentrically circular tracks. Given the same rotational speed of an HDD, the outer tracks would have higher data throughput than the inner ones. As such, those regular patterns might indicate that the HDD heads sequentially shuttle between outer and inner tracks when consecutively writing/reading block data during the experiments.

\begin{figure}[!t]
  \centering
  \includegraphics[width= 7.4cm]{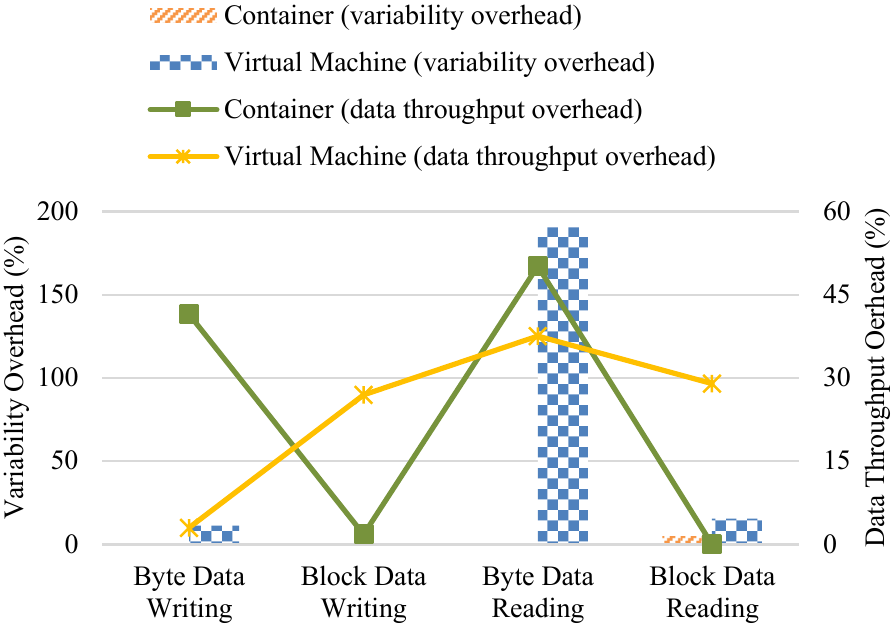}
  \caption{Storage data throughput and its variability overhead of a standalone Docker container vs.~VM (using the benchmark Bonnie++).}
  \label{fig:storageOverhead} 
\end{figure}

The second surprising observation is that, unlike most cases in which the VM has the worst performance, the container seems significantly poor at accessing the byte size of data, although its performance variability is clearly the smallest. We further calculate the storage performance overhead to deliver more specific comparison between the container and the VM, and draw the results into Figure \ref{fig:storageOverhead}. Note that, in the case when the container's/VM's variability is smaller than the physical machine's, we directly set the corresponding variability overhead to zero rather than allowing any performance overhead to be negative. Then, the bars in the chart indicate that the storage variability overheads of both virtualization technologies are nearly negligible except for reading byte-size data on the VM (up to nearly $200\%$). Although the storage driver is a known bottleneck for a container's internal disk I/O, it is still surprising that the container brings around $40\%$ to $50\%$ data throughput overhead when performing disk operations on a byte-by-byte basis. In other words, the container solution might not be suitable for the Cloud applications that have frequent and unpredictable intermediate data generation and consumption. On the contrary, there is relatively trivial performance loss in VM's byte data writing. However, the VM has roughly $30\%$ data throughput overhead in other disk I/O scenarios, whereas the container barely incurs overhead when reading/writing large size of data.

Our third observation is that, the storage performance overhead of different virtualization technologies can also be reflected through the total number of the iterative Bonnie++ trials. As pointed by the maximum x-axis scale in Figure \ref{fig:bonnieRun}, the physical machine, the container and the VM can respectively finish 150, 147 and 101 rounds of disk tests during 24 hours. Given this information, we estimate the container's and the VM's storage performance overhead to be $2\%~(=|147-150|/150)$ and $32.67\%~(=|101-150|/150)$ respectively.

\section{Conclusion}
\label{sec:conclusion}


Following the performance evaluation methodology DoKnowMe, we draw conclusions mainly by answering the predefined requirement questions. Driven by RQ1 and RQ2, our evaluation result largely confirms the aforementioned qualitative discussions: The container's average performance is generally better than the VM's and is even comparable to that of the physical machine with regarding to many features. Specifically, the container has less than $4\%$ performance overhead in terms of communication data throughput, computation latency, memory data throughput and storage data throughput. Nevertheless, the container-based virtualization could hit a bottleneck of storage transaction speed, with the overhead up to $50\%$. Note that, as mentioned previously, we interpret the byte-size data throughput into storage transaction speed, because each byte essentially calls a disk transaction here. In contrast, although the VM delivers the worst performance in most cases, it could perform as well as the physical machine when solving the N-Queens problem or writing small-size data to the disk.

Driven by RQ3 and RQ4, we find that the performance loss resulting from virtualizations is more visible in the performance variability. For example, the container's variability overhead could reach as high as over $500\%$ with respect to the Fibonacci calculation and the memory Triad operation. Similarly, although the container generally shows less performance variability than the VM, there are still exceptional cases: The container has the largest performance variation in the job of computing Fourier transform, whereas even the VM's performance variability is not worse than the physical machine's when running CryptoHash, N-Queens, and Raytracing jobs.

Overall, our work reveals that the performance overheads of these two virtualization technologies could vary not only on a feature-by-feature basis but also on a job-to-job basis. Although the container-based solution is undoubtedly lightweight, the hypervisor-based technology does not come with higher performance overhead in every case. Based on such a fundamental evaluation study, we will gradually apply Docker containers to different real-world applications in the coming future. The application-oriented practices will also be replicated in the hypervisor-based virtual environment for further comparison case studies.

\section*{Acknowledgment}
This work is supported by the Swedish Research Council
(VR) for the project ``Cloud Control", and through the
LCCC Linnaeus and ELLIIT Excellence Centers.


\bibliographystyle{IEEEtran}
\bibliography{AINA2017_Ref}

\begin{thebibliography}{10}
\providecommand{\url}[1]{#1}
\csname url@samestyle\endcsname
\providecommand{\newblock}{\relax}
\providecommand{\bibinfo}[2]{#2}
\providecommand{\BIBentrySTDinterwordspacing}{\spaceskip=0pt\relax}
\providecommand{\BIBentryALTinterwordstretchfactor}{4}
\providecommand{\BIBentryALTinterwordspacing}{\spaceskip=\fontdimen2\font plus
\BIBentryALTinterwordstretchfactor\fontdimen3\font minus
  \fontdimen4\font\relax}
\providecommand{\BIBforeignlanguage}[2]{{%
\expandafter\ifx\csname l@#1\endcsname\relax
\typeout{** WARNING: IEEEtran.bst: No hyphenation pattern has been}%
\typeout{** loaded for the language `#1'. Using the pattern for}%
\typeout{** the default language instead.}%
\else
\language=\csname l@#1\endcsname
\fi
#2}}
\providecommand{\BIBdecl}{\relax}
\BIBdecl

\bibitem{Pahl_2015}
C.~Pahl, ``{Containerization and the PaaS Cloud},'' \emph{IEEE Cloud Comput.},
  vol.~2, no.~3, pp. 24--31, May/Jun. 2014.

\bibitem{Walters_Chaudhary_2008}
J.~P. Walters, V.~Chaudhary, M.~Cha, S.~G. Jr., and S.~Gallo, ``A comparison of
  virtualization technologies for {HPC},'' in \emph{Proc. 22nd Int. Conf. Adv.
  Inf. Networking Appl. (AINA 2008)}.\hskip 1em plus 0.5em minus 0.4em\relax
  Okinawa, Japan: IEEE Computer Society, 25-28 Mar. 2008, pp. 861--868.

\bibitem{Feitelson_2007}
D.~G. Feitelson, ``Experimental computer science,'' \emph{Commun. ACM},
  vol.~50, no.~11, pp. 24--26, Nov. 2007.

\bibitem{Li_OBrien_DoKnowMe}
Z.~Li, L.~O'Brien, and M.~Kihl, ``{DoKnowMe: Towards} a domain knowledge-driven
  methodology for performance evaluation,'' \emph{ACM SIGMETRICS Perform. Eval.
  Rev.}, vol.~43, no.~4, pp. 23--32, Mar. 2016.

\bibitem{Merkel_2014}
D.~Merkel, ``Docker: {Lightweight Linux} containers for consistent development
  and deployment,'' \emph{Linux J.}, vol. 239, pp. 76--91, Mar. 2014.

\bibitem{Xu_Yu_2014}
X.~Xu, H.~Yu, and X.~Pei, ``A novel resource scheduling approach in container
  based clouds,'' in \emph{Proc. 17th IEEE Int. Conf. Comput. Sci. Eng. (CSE
  2014)}.\hskip 1em plus 0.5em minus 0.4em\relax Chengdu, China: IEEE Computer
  Society, 19-21 Dec. 2014, pp. 257--264.

\bibitem{Bernstein_2014}
D.~Bernstein, ``{Containers and Cloud: From LXC to Docker to KuBernetes},''
  \emph{IEEE Cloud Comput.}, vol.~1, no.~3, pp. 81--84, Sept. 2014.

\bibitem{Xavier_Neves_2014}
M.~G. Xavier, M.~V. Neves, and C.~A. F.~D. Rose, ``A performance comparison of
  container-based virtualization systems for {MapReduce} clusters,'' in
  \emph{Proc. 22nd Euromicro Int. Conf. Parallel Distrib. Network-Based
  Process. (PDP 2014)}.\hskip 1em plus 0.5em minus 0.4em\relax Turin, Italy:
  IEEE Press, 12-14 Feb. 2014, pp. 299--306.

\bibitem{Anderson_2015}
C.~Anderson, ``Docker,'' \emph{IEEE Software}, vol.~32, no.~3, pp. 102--105,
  May/Jun. 2015.

\bibitem{Banerjee_2014}
T.~Banerjee, ``Understanding the key differences between {LXC and Docker},''
  \url{https://www.flockport.com/lxc-vs-docker/}, Aug. 2014.

\bibitem{Che_Shi_2010}
J.~Che, C.~Shi, Y.~Yu, and W.~Lin, ``A synthetical performance evaluation of
  {OpenVZ, Xen and KVM},'' in \emph{Proc. 2010 IEEE Asia-Pacific Serv. Comput.
  Conf. (APSCC 2010)}.\hskip 1em plus 0.5em minus 0.4em\relax Hangzhou, China:
  IEEE Computer Society, 6-10 Dec. 2010, pp. 587--594.

\bibitem{Strauss_2013}
D.~Strauss, ``Containers - not virtual machines - are the future {Cloud},''
  \emph{Linux J.}, vol. 228, pp. 118--123, Apr. 2013.

\bibitem{Adufu_Choi_2015}
T.~Adufu, J.~Choi, and Y.~Kim, ``Is container-based technology a winner for
  high performance scientific applications?'' in \emph{Proc. 17th Asia-Pacific
  Network Oper. Manage. Symp. (APNOMS 2015)}.\hskip 1em plus 0.5em minus
  0.4em\relax Busan, Korea: IEEE Press, 19-21 Aug. 2015, pp. 507--510.

\bibitem{Joy_2015}
A.~M. Joy, ``Performance comparison between {Linux} containers and virtual
  machines,'' in \emph{Proc. 2015 Int. Conf. Adv. Comput. Eng. Appl. (ICACEA
  2015)}.\hskip 1em plus 0.5em minus 0.4em\relax Ghaziabad, India: IEEE Press,
  14-15 Feb. 2015, pp. 507--510.

\bibitem{Seo_Hwang_2014}
K.-T. Seo, H.-S. Hwang, I.-Y. Moon, O.-Y. Kwon, and B.-J. Kim, ``Performance
  comparison analysis of {Linux} container and virtual machine for building
  {Cloud},'' \emph{Adv. Sci. Technol. Lett.}, vol.~66, pp. 105--111, Dec. 2014.

\bibitem{Felter_Ferreira_2015}
W.~Felter, A.~Ferreira, R.~Rajamony, and J.~Rubio, ``An updated performance
  comparison of virtual machines and {Linux} containers,'' in \emph{Proc. 2015
  IEEE Int. Symp. Perform. Anal. Syst. Software (ISPASS 2015)}.\hskip 1em plus
  0.5em minus 0.4em\relax Philadelphia, PA, USA: IEEE Press, 29-31 Mar. 2015,
  pp. 171--172.

\bibitem{Dua_Raja_2015}
R.~Dua, A.~R. Raja, and D.~Kakadia, ``Virtualization vs containerization to
  support {PaaS},'' in \emph{Proc. 2014 IEEE Int. Conf. Cloud Eng. (IC2E
  2015)}.\hskip 1em plus 0.5em minus 0.4em\relax Boston, Massachusetts, USA:
  IEEE Computer Society, 10-14 Mar. 2014, pp. 610--614.

\bibitem{Piraghaj_Dastjerdi_2015}
S.~F. Piraghaj, A.~V. Dastjerdi, R.~N. Calheiros, and R.~Buyya, ``Efficient
  virtual machine sizing for hosting containers as a service,'' in \emph{Proc.
  11th World Congr. Serv. (SERVICES 2015)}.\hskip 1em plus 0.5em minus
  0.4em\relax New York, USA: IEEE Computer Society, 27 Jun.-2 Jul. 2015, pp.
  31--38.

\bibitem{Blackburn_McKinley_2008}
S.~M. Blackburn, K.~S. McKinley, R.~Garner, C.~Hoffmann, A.~M. Khan,
  R.~Bentzur, A.~Diwan, D.~Feinberg, D.~Frampton, S.~Z. G. M. H. A. H. M. J.~H.
  Lee, J.~E.~B. Moss, A.~Phansalkar, D.~Stefanovik, T.~VanDrunen, D.~von
  Dincklage, and B.~Wiedermann, ``Wake up and smell the coffee: {Evaluation}
  methodology for the 21st century,'' \emph{Commun. ACM}, vol.~51, no.~8, pp.
  83--89, Aug. 2008.

\bibitem{Iosup_Yigitbasi_2011}
A.~Iosup, N.~Yigitbasi, and D.~Epema, ``On the performance variability of
  production {Cloud} services,'' in \emph{Proc. 11th IEEE/ACM Int. Symp.
  Cluster Cloud Grid Comput. (CCGrid 2011)}.\hskip 1em plus 0.5em minus
  0.4em\relax Newport Beach, CA, USA: IEEE Computer Society, 23-26 May 2011,
  pp. 104--113.

\bibitem{Li_OBrien_2012_taxonomy}
Z.~Li, L.~O'Brien, R.~Cai, and H.~Zhang, ``Towards a taxonomy of performance
  evaluation of commercial {Cloud} services,'' in \emph{Proc. 5th Int. Conf.
  Cloud Comput. (IEEE CLOUD 2012)}.\hskip 1em plus 0.5em minus 0.4em\relax
  Honolulu, Hawaii, USA: IEEE Computer Society, 24-29 Jun. 2012, pp. 344--351.

\bibitem{Li_OBrien_2012_factor}
Z.~Li, L.~O'Brien, H.~Zhang, and R.~Cai, ``A factor framework for experimental
  design for performance evaluation of commercial {Cloud} services,'' in
  \emph{Proc. 4th Int. Conf. Cloud Comput. Technol. Sci. (CloudCom
  2012)}.\hskip 1em plus 0.5em minus 0.4em\relax Taipei, Taiwan: IEEE Computer
  Society, 3-6 Dec. 2012, pp. 169--176.

\bibitem{StackOverflow_2013}
StackOverflow, ``What is the relationship between the docker host {OS} and the
  container base image {OS}?''
  \url{http://stackoverflow.com/questions/18786209/what-is-the-relationship-between-the-docker-host-os-and-the-container-base-image},
  Sept. 2013.

\bibitem{Reddit_2015}
Reddit, ``Do {I} need to use an {OS} base image in my {Dockerfile} or will it
  default to the host {OS}?''
  \url{https://www.reddit.com/r/docker/comments/2teskf/do_i_need_to_use_an_os_base_image_in_my/},
  Jan. 2015.

\bibitem{Li_OBrien_2014_blueprint}
Z.~Li, L.~O'Brien, H.~Zhang, and R.~Cai, ``On the conceptualization of
  performance evaluation of {IaaS} services,'' \emph{IEEE Trans. Serv.
  Comput.}, vol.~7, no.~4, pp. 628--641, Oct.-Dec. 2014.

\end{thebibliography}

\end{document}